\documentclass[reprint,amsmath,amssymb,aps,prl,longbibliography]{revtex4-1}

\usepackage[utf8]{inputenc}
\usepackage[T1]{fontenc}
\usepackage{lmodern}
\usepackage[english]{babel}
\usepackage[autostyle=true]{csquotes}
\usepackage[colorlinks=true,  urlcolor=blue,  linkcolor=blue,  citecolor=blue]{hyperref}
\usepackage[dvipsnames]{xcolor}
\usepackage{siunitx}
\usepackage{graphicx}

\newcommand{\bs}{\boldsymbol}

\newcommand{\EJD}{\epsilon_q}
\newcommand{\mum}{\rm \si{\micro\meter}}
\newcommand{\rhoe}{\rho_{\rm eq}}
\newcommand{\ket}[1] {\left\vert #1 \right\rangle}

\begin{document}

\title{Superfluid Fraction in an Interacting  Spatially Modulated Bose-Einstein Condensate} 

\author{G. Chauveau, C. Maury, F. Rabec, C. Heintze, G. Brochier, S. Nascimbene, J. Dalibard, J. Beugnon}

\email{beugnon@lkb.ens.fr}

\affiliation{Laboratoire Kastler Brossel,  Coll\`ege de France, CNRS, ENS-PSL University, Sorbonne Universit\'e, 11 Place Marcelin Berthelot, 75005 Paris, France}

\author{S.M. Roccuzzo and S. Stringari}
\email{sandro.stringari@unitn.it}
\affiliation{Pitaevskii BEC Center, CNR-INO and Dipartimento di Fisica, Università di Trento, I-38123 Trento, Italy}
\affiliation{Trento Institute for Fundamental Physics and Applications, INFN, Italy}

\date{\today}

\begin{abstract}
At zero temperature, a Galilean-invariant Bose fluid is expected to be fully superfluid. Here we investigate theoretically and experimentally the quenching of the superfluid density of a dilute Bose-Einstein condensate due to the breaking of translational (and thus Galilean) invariance by an external 1D periodic potential. Both Leggett's bound fixed by the knowledge of the total density and the anisotropy of the sound velocity provide a consistent determination of the superfluid fraction. The use of a large-period lattice emphasizes the important role of two-body interactions on superfluidity.
\end{abstract}

\maketitle

Superfluidity is a unique state of matter exhibited by quantum many-body systems in special conditions of temperature and interactions. It is characterized by the absence of viscosity and by many other peculiar phenomena, like the occurrence of quantized vortices, the reduction of the moment of inertia, the propagation of second sound at finite temperature, and Josephson effects. Superfluidity was first discovered in liquid helium \cite{Kapitza1938,Allen1938}. More recently, an impressive amount of scientific activity has concerned the superfluid behavior of ultracold atomic gases (for a review see Refs.\,\cite{BecBook2016,Pethick08, Leggett06}).

A key quantity characterizing superfluidity is the fraction of the total density, the so-called superfluid fraction, which determines superfluid transport phenomena. According to Landau's theory  of superfluidity \cite{Landau41}, at nonzero temperature the superfluid density does not coincide with the total density. The  thermal occupation of elementary excitations  provides the normal (non-superfluid) component  responsible, for example, for the non-vanishing moment of inertia and the propagation of second sound \cite{BecBook2016}. The measurement of second sound velocity  provides unique information on the temperature dependence of the superfluid density, in both liquid helium \cite{peshkov1944second} and quantum gases \cite{Sidorenkov13,Christodoulou20}.

At zero temperature, superfluid and total densities still do not always coincide as illustrated by the celebrated superfluid to Mott-insulator transition \cite{Fisher89,Jaksch98,Greiner02}. Even in the mean-field regime relevant for the present work, consistent with the applicability of Gross-Pitaevskii theory, quenching of the superfluid density can occur when translation or Galilean invariances are broken, resulting in 
important consequences on the excitation spectrum.   Such effects have been already pointed out theoretically in the presence of disorder \cite{giorgini1994effects,gaul2009speed}, external periodic potentials \cite{javanainen1999phonon,kramer2002macroscopic, smerzi2002dynamical,machholm2003band,taylor2003bogoliubov,watanabe2008equation},  supersolidity \cite{saccani2012excitation,macri2013elementary}  and spin-orbit coupling \cite{li2013superstripes,martone2012anisotropic,zhang2016superfluid}. Experimentally, the effects of the quenched superfluidity on the collective frequencies of a harmonically trapped gas were investigated in Ref.\,\cite{cataliotti2001josephson}. A similar situation emerges in astrophysics in the context of neutron stars where the periodic lattice of nuclei  influences the superfluid density in the inner crust  \cite{Chamel12, Watanabe17}.

Here we provide a combined theoretical and experimental investigation of the reduction of the superfluid fraction caused by the presence of a periodic potential
in a weakly interacting Bose-Einstein condensate (BEC) confined in a box.  We determine the superfluid density employing Leggett's result \cite{Leggett1970,Leggett1998}, which is based on the knowledge of the {\it{in situ}} modulated total density profile $\rho(\bs r)$, experimentally available thanks to the use of a large-period lattice. We also present an independent measurement of the superfluid fraction taht exploits the anisotropic character of the sound velocity. 


\paragraph{Superfluid fraction in a modulated potential.}

We consider a two-dimensional (2D) weakly interacting BEC confined in a box of size $L\times L$, in the presence of the one-dimensional (1D) spatially periodic potential 
$V(x) = V_0 \cos\left(qx\right)$. This potential brings two energy scales to the problem, 
its amplitude $V_0$ and the ``recoil energy'' $\EJD =\hbar^2q^2/2m$, where $m$ is the mass of an atom. Atomic interactions provide the third energy scale relevant for the problem. They are conveniently characterized by the chemical potential $\mu_0=g\rho_0$ of a uniform condensate with a density equal to the average value $ \rho_0=\langle \rho(x) \rangle$, where $\rho(x)$ is the density profile and the average is calculated over one period of the potential $V(x)$. The  interaction coupling constant between atoms $g=4\pi\hbar^2 a_s/m$ is fixed by the $s$-wave scattering length $a_s$. 

In such a configuration, the superfluid fraction is an anisotropic rank-2 tensor with eigenaxes $x,y$ and with diagonal elements denoted $f_{s,\alpha}$ in the following ($\alpha=x,y$). They can be calculated by applying the perturbation $-v\hat P_\alpha$ to the system, where $\hat P_\alpha=\sum_{j=1}^N \hat p_{j,\alpha}$ is  the momentum operator along the axis $\alpha$ and $N$ the number of particles. This corresponds to working in the frame moving with velocity $v$ with respect to the laboratory frame. Only the normal part reacts to the perturbation, so that, by calculating the average momentum $\langle \hat P_\alpha\rangle$ and imposing periodic boundary conditions, one  accesses to the superfluid  fraction along the axis $\alpha$
\begin{equation}
f_{s,\alpha}=1- \lim_{v\to 0} \frac{\langle \hat P_\alpha\rangle}{Nmv} \; .
\label{movingframe}
\end{equation} 
A similar  procedure, applied to the case of a rotating configuration, employing the angular momentum operator rather than  the linear momentum operator, gives access to the moment of inertia, whose deviation from the rigid value provides direct evidence of superfluid effects \cite{stringari1996moment}.

In the presence of the periodic potential $V(x)$, the motion of the fluid is slowed down along the $x$ direction, reflecting the quenching of the superfluid density along this direction: $\rho_{s,x}<\rho_0$. The superfluid density evaluated along the transverse $y$ direction is instead not modified: $\rho_{s,y}=\rho_0$. The Gross-Pitaevskii equation (GPE) describing the weakly interacting BEC,  solved in  the frame moving with velocity $v$, i.e. subject to the constraint
 $-vP_x$,  yields, according to the definition (\ref{movingframe}), the result  \cite{REFSM}
\begin{equation}
    f_{s,x} = \frac{\rho_{s,x}}{\rho_0} 
    =\frac{1}{\langle \rho(x)\rangle\ \langle \frac{1}{\rho(x)}\rangle}.
    \label{leggett}
\end{equation}
According to the seminal work by Leggett \cite{Leggett1970,Leggett1998},  the right-hand side of (\ref{leggett}) provides generally  an upper bound to the superfluid density. Remarkably, the bound reduces to an identity in the case of a weakly interacting BEC subject to a 1D periodic potential \footnote{Result (\ref{leggett}) for the superfluid density does not hold only in  the  presence of  periodic  potentials of the form $V_0 \cos\left(qx\right)$. For example, it provides the  Josephson energy in the presence of a  junction separating two BECs \cite{zapata1998josephson}}.

\paragraph{Effective mass and sound propagation.}
Result (\ref{leggett}) may be surprising because it relates a transport property (the superfluid density) to a static quantity (the equilibrium density profile). The concept of effective mass, commonly used in the context of interacting Bose \cite{kramer2002macroscopic,zwerger2003mott,taylor2003bogoliubov}  and Fermi gases  \cite{watanabe2008equation} placed in a periodic potential, elucidates this relation. In the present case, the superfluid fraction of the BEC, defined according to  (\ref{leggett}),  
exactly coincides with the ratio 
\begin{equation}
\frac{m}{m_x^*}=f_{s,x}\;,
\label{m*rhos}
\end{equation}
 where the effective  mass $m_x^*$ fixes the curvature of the energy band along the $x$ direction for small values of the quasi-momentum \cite{REFSM}.

The relation (\ref{m*rhos}) between the effective mass and $f_{s,\alpha}$ illustrates the crucial role of the superfluid density in the propagation of sound. The hydrodynamic formalism of superfluids indeed provides the following expression for the velocity of a sound wave propagating along the $x$  direction in the presence of $V(x)$ \cite{kramer2002macroscopic,BecBook2016,bensoussan2011asymptotic,cataliotti2001josephson}
\begin{equation}
c^2_x= \frac{1}{m_x^*\kappa} ={f_{s,x}}\;\frac{1}{m\kappa}\; ,
\label{cx}
\end{equation}
where   $\kappa=\left[\rho_0\,\partial_{\rho_0}\mu(\rho_0)\right]^{-1}$ is the  compressibility of the gas. 
\begin{figure}[t]
\centering
\includegraphics[width=7cm]{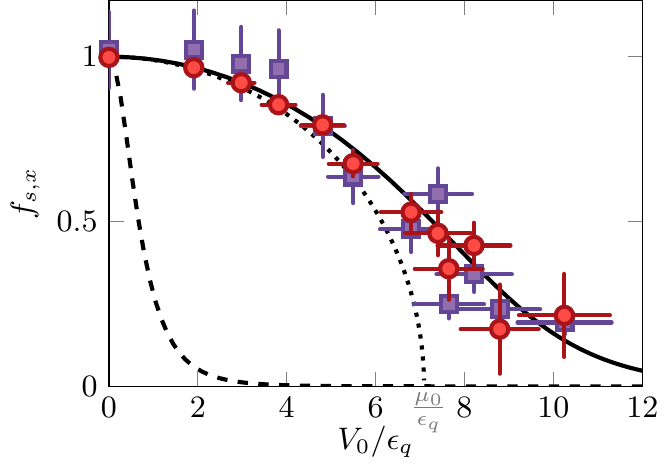}
\caption{Solid line: Superfluid fraction calculated for a $^{87}$Rb condensate by injecting the  numerical solution of the GPE inside Leggett's formula (\ref{leggett}) for $\rho_0=60\,\mum^{-2}$, $mg/\hbar^2=0.15$ and $2\pi/q=3.93\,\mum$, corresponding to $\mu_0/\epsilon_q=7.0$. Dotted line: LDA result. 
Red circles: Experimental results obtained using Leggett's formula. Violet squares: Experimental results obtained from speeds of sound. The dashed line shows the prediction in the opposite  regime, $\mu_0 \ll \epsilon_q$, of a very weakly-interacting system. In all figures, the error bar represents the statistical uncertainties of the measurements.}\label{fig:fsmeasured}
\end{figure}
The value of the sound velocity propagating along the transverse $y$ direction is
\begin{equation}
c^2_y=\frac{1}{m \kappa} \; ,
\label{cy}
\end{equation}
the effective mass $m_y^*$ being equal to the bare mass in this case. 
The ratio between Eqs. \ref{cx} and \ref{cy} then provides 
\begin{equation}
f_{s,x}= \frac{c^2_x}{c^2_y}.
\label{fscxcy}
\end{equation}
The superfluid fraction can thus be determined either through the explicit knowledge of the equilibrium density profile via \eqref{leggett} or through the measurement of the ratio \eqref{fscxcy}.

\paragraph{Limiting cases.}

Results \eqref{leggett} and \eqref{fscxcy} hold for any values of the dimensionless parameters $V_0/\mu_0$ and $\EJD /\mu_0$, as long as the description of the $T=0$ Bose gas by a macroscopic wave function is valid, i.e., as long as quantum phase fluctuations between neighboring sites (a precursor of  the superfluid to Mott-insulator transition) can be ignored.
 We now examine some limiting cases where $\rho(x)$ and $f_{s,x}$ take a simple expression.


We start with the very weakly-interacting regime where $\mu_0 \ll \epsilon_q$. In this case, the GPE  approaches  the Schr\"odinger equation for a single particle subject to the periodic potential $V(x)$. In this regime, the identity $m^*/m=\langle \rho\rangle\, \langle 1/\rho\rangle $  was already noticed in Ref.\,\cite{Leggett2022}. 

The opposite case  $\EJD  \ll \mu_0 $ is described by the Local Density Approximation (LDA). The validity condition of the LDA is equivalent to imposing that the period of the potential $2\pi/q$ be much larger than the healing length $\hbar/ \sqrt{2 mg\rho_0}$. In the LDA, the equilibrium density becomes $\rho^{\rm (LDA)}(x)=\rho_0 - \rho_1 \cos (qx)$ with $\rho_0=\mu_0/g$ and $\rho_1=V_0/g$. When injected into (\ref{leggett}), this gives 
\begin{equation}
f^{\rm (LDA)}_{s,x}=\left(1-\frac{\rho_1^2}{\rho_0^2} \right)^{1/2}=\left(1-\frac{V_0^2}{\mu_0^2} \right)^{1/2}.
\end{equation}
For values $V_0 > \mu_0$, the LDA density vanishes in a finite region within each period of $V(x)$, with the consequent vanishing of the superfluid fraction, (dotted line in Fig.\,\ref{fig:fsmeasured}). 

We now consider the third limiting case of small $V_0$, which  can be addressed using the  formalism of the static density response function. An expansion of the solution of the GPE in powers of $V_0$ yields the amplitude of the first Fourier component of $\rho(x)$
\begin{equation}
\frac{\rho_1}{\rho_0}= \frac{2V_0}{2\mu_0+\EJD } +{\cal O}(V_0^3).
\label{eq:amplitude}
\end{equation}
Using (\ref{leggett}), one finds the superfluid fraction \cite{taylor2003bogoliubov}
\begin{equation}
f_{s,x}= 1- \frac{2V^2_0}{(2\mu_0+\EJD )^2} +{\cal O}(V_0^4) \; ,
\label{fschi}
\end{equation}
confirming the LDA result given above when we take the limit $\EJD /\mu_0\to 0$. Note that (\ref{eq:amplitude},\ref{fschi}) also hold in the opposite limit of large $\EJD /\mu_0 $, where the superfluid density  no longer depends on the  interaction. 

The expansion of the solution of the GPE in powers of $V_0$ also provides the  compressibility 
\begin{equation}
\kappa = \mu_0^{-1}\left[1-\frac{2V_0^2 \EJD }{(2\mu_0+\EJD )^3}\right] +{\cal O}(V_0^4)
\label{kappa} \; ,
\end{equation}
showing that, different from the expansion (\ref{fschi}) for the superfluid density, the $V_0^2$ correction to the compressibility vanishes in the LDA limit $\EJD /\mu_0 \to 0$. In this limit, we thus predict from \ref{cx} and \ref{cy} that, at order 2 in $V_0$, the speed of sound $c_y$ in the direction perpendicular to the lattice will not be affected by the presence of the lattice, whereas $c_x$ will be reduced by an amount directly related to $f_{s,x}$.

\paragraph{The ideal gas limit.}
The addition of a lattice on a  Bose gas sheds interesting light on the controversial question of the possible superfluidity in the ideal case. The fact that the Landau criterion is not satisfied points to a non-superfluid character of the ideal gas, while the approach based on twisted boundary conditions leads to $f_s=1$ for this system. To remove this ambiguity, we take a gas with chemical potential $\mu_0$ placed in a lattice of large spatial period ($\epsilon_q\ll V_0$) and consider the two limits (i) $V_0 \to 0$  and (ii) $\mu_0\to 0$. The order in which these limits are taken is crucial. If we take limit (i) first (\textsl{i.e.} $\epsilon_q\ll V_0\ll  \mu_0$) and then limit (ii), we find $f_s=1$, see \eqref{fschi}. Conversely, taking first the limit (ii) (\textsl{i.e.} $\epsilon_q, \mu_0\ll  V_0 $) leads to $f_s\approx 0$ (see dashed line in Fig.\,1). 
In our opinion, the latter approach is more relevant as it implicitly takes into account the residual (possibly disordered) modulated potentials acting on the gas.


\paragraph{Experimental setup.}

We now describe the experimental determination of the superfluid fraction of a planar BEC subjected to a sinusoidal potential along $x$. The setup has  been detailed in Refs.\,\cite{Ville17,Ville18}. We start from a single quasi-2D Bose gas of $^{87}$Rb atoms confined in an optical dipole trap made of a combination of repulsive laser beams  at a wavelength $\lambda=532\,$nm. We load all atoms around a  single node of an optical lattice, which provides a strong confinement along  the vertical  direction $z$. It leads to an approximate harmonic confinement of frequency $\omega_z/2\pi\approx3.7$\,kHz. The associated characteristic length, $\ell_z=\sqrt{\hbar/m\omega_z}\approx180$\,nm, is large compared to the s-wave scattering length $a_s=5.3\,$nm, which leads to a quasi-2D regime where collisions keep their three-dimensional (3D) character. The effective 2D coupling constant describing the interactions in the cloud is  $m g/\hbar^2= \sqrt{8\pi} a_s /\ell_z\approx 0.15$. The 2D character of our gas is not crucial for this  experiment, which could also be performed in a 3D box-like potential \cite{Gaunt13}.

The in-plane confinement is created by spatially-shaped laser beams. A first beam creates a square box potential of size $L=40\,\mum$. A second  beam imposes the sinusoidal potential modulated along the $x$ axis with a tunable amplitude $V_0$ from 0 to 80\,nK \cite{Zou21,REFSM}. The lattice period $d=3.93(4)\,\mum$ and the average 2D density $\rho_0=60(3)\,\mum^{-2}$ are fixed. This corresponds to $\mu_0/k_B\approx50\,$nK and $\epsilon_q/k_B=7.1\,$nK. The temperature of the gas is below the lowest measurable value in our setup, i.e.  $<20\,$nK.

\begin{figure}[t!]
    \centering
\includegraphics[width=8cm]{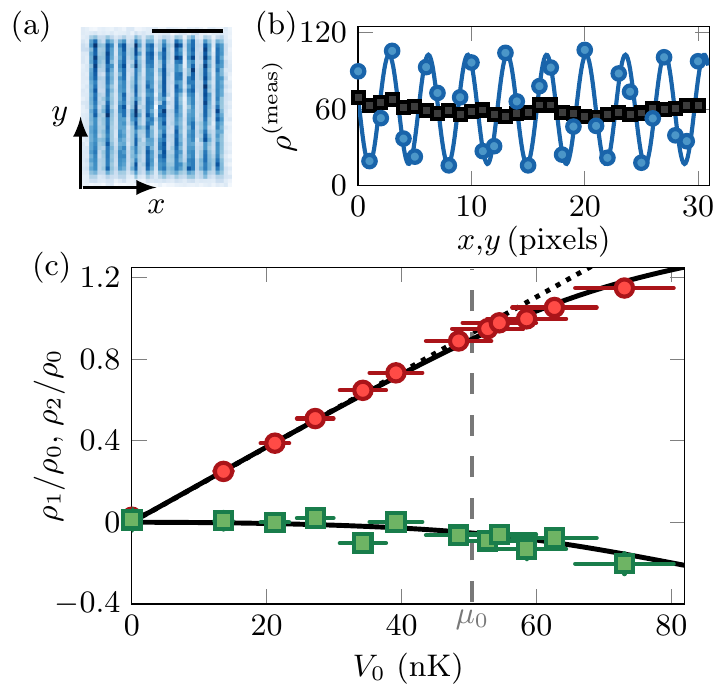}
\caption{(a) \textit{In situ} absorption image of the 2D gas modulated with a lattice along $x$ of period $3.93\,\mum$ and amplitude $V_0= 54(5)$\,nK. The length of the scale bar is 20\,$\mum$. (b) Density profile integrated along $x$ (squares) and $y$ (circles) and the fit to a sinusoidal modulation (solid blue line) for $\rho^{\rm (meas)}(x)$. The pixel size is 1.15\,$\mum$. (c) Fourier components of the density modulation  $\rho_n/\rho_0$ versus the lattice depth $V_0$ for $n=1$\,(circles) and $n=2$\,(squares).
Solid lines represent the corresponding predictions from the GPE. The dotted line is the weak lattice limit of \eqref{eq:amplitude}.}
    \label{fig:densitymodulation}
\end{figure}

\begin{figure}[t]
\centering
\includegraphics[width=7cm]{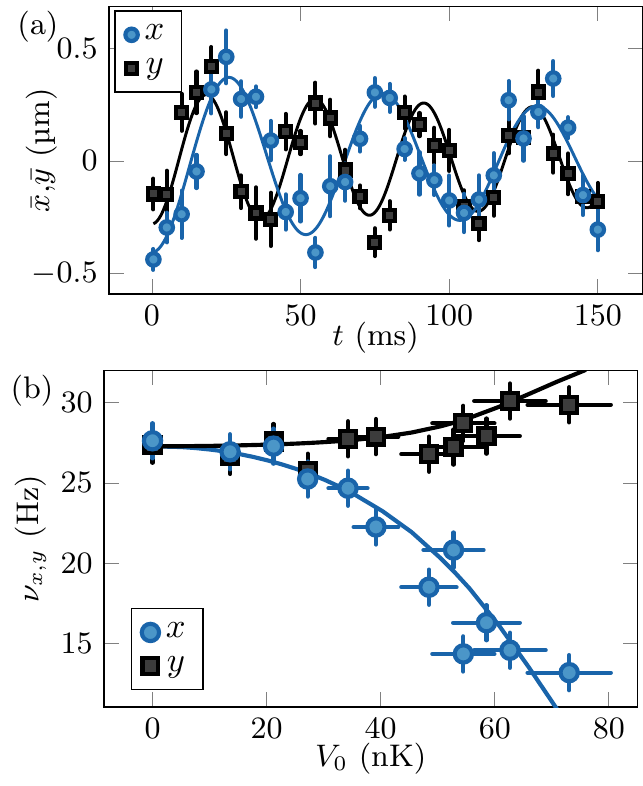}
\caption{Speed of sound measurement. (a) Center of mass position of the cloud after excitation as a function of time for an excitation along or perpendicular to the lattice for $V_{0}/k_B=41(4)$ nK. The solid lines are sinusoidal fits to the data giving $\nu_x=19(1)\,$Hz and $\nu_y=27(1)\,$Hz.  (b) Extracted speed of sound along $y$ (black squares) and $x$ (blue circles) axes for different lattice depths. The solid lines are the prediction from the GPE.}\label{fig:sound}
\end{figure}


\paragraph{Superfluid fraction from Leggett's formula.}

To use Leggett's result (\ref{leggett}), we measure the \textit{in situ} 2D density profile $\rho^{\rm (meas)}(x,y)$ in the presence of the lattice using absorption imaging, see Fig.\,\ref{fig:densitymodulation}(a). We integrate it along $y$ to obtain the 1D profile $\rho^{\rm (meas)}(x)$ (Fig.\,\ref{fig:densitymodulation}(b)). 
For an ideal imaging system \footnote{We use a weakly-saturating imaging beam and partial transfer imaging to image low density clouds and stay in the linear absorption regime.}, $\rho^{\rm (meas)}(x)=\rho(x)$ but finite optical resolution alters this relation and has to be included in the analysis. The expected density distribution can be expanded in Fourier series $\rho_0-\sum_{n>0}\rho_n\cos(nqx)$, where the role of higher harmonics becomes increasingly important for large $V_0$. We model our optical resolution by multiplicative coefficients $\beta_n<1$: $\rho^{\rm (meas)}(x)=\rho_0-\sum_{n>0}\beta_n\rho_n\cos(nqx)$.

We calibrate the first coefficients $\beta_n$ by  studying the density response to a lattice of wave number $q'=nq$
for low lattice depths. In this case, the density modulation $\rho(x)$ is dominated by its first harmonic and we fit the measured profiles to a sinusoidal function whose amplitude is adjusted to prediction (\ref{eq:amplitude}). This adjustment provides $\beta_1=0.73(2)$ and $\beta_2=0.27(6)$, while the values of the coefficients $n\geq3$ are below our experimental detectivity.

We show in Fig.\,\ref{fig:densitymodulation}(c) the values of $\rho_n=\rho_n^{(\rm meas)}/\beta_n$ for $n=1,2$. Both measurements are in good agreement with the predictions of the GPE (solid lines) over all the explored range of values of $V_0$. From this measurement and restricting $\rho(x)$ to its two first Fourier components, we calculate Leggett's formula \eqref{leggett} and we plot the result as circles in Fig.\,\ref{fig:fsmeasured}. We discuss in Ref.\,\cite{REFSM} the effect of the truncation of the Fourier series on the solution of the GPE and confirm that, in our case, restricting to the first two harmonics already gives a good estimate of $f_{s,x}$.


\paragraph{Superfluid fraction from speed of sound.}
We determine the speeds of sound along $x$ and $y$ by studying the response of the cloud to an external perturbation of its density. Here, the perturbation consists  adding, during the preparation of the cloud, a  weak linear magnetic potential along $x$ or $y$, of amplitude $\approx$ 0.1\,nK/$\mum$. At time $t=0$, we abruptly switch off this potential and measure the evolution of the center of mass of the cloud, see Fig.\,\ref{fig:sound}(a). For a perturbation along $x$, we observe a smaller frequency than the one obtained for the same excitation along the $y$ axis, a clear signature of the modification of the superfluid transport properties due to the presence of  the lattice. 

From the frequency $\nu_{x,y}$ of the fitted oscillations, we determine the speed of sound through the relation $c_{x,y}=2L \nu_{x,y}$ valid when the lattice period and the healing length are both much smaller than the phonon wavelength, equal here to $2L$. We show our measurements in Fig.\,\ref{fig:sound}(b) as a function of the lattice depth.  For a perturbation along $y$, we observe a small increase of $c_y$ with $V_0$, which  can be attributed to the modification of the compressibility of the modulated gas with respect to the uniform case (see Eqs. \ref{cy} and \ref{kappa}). Along the axis of the lattice, we note a strong decrease of the speed of sound with the amplitude of the modulating potential,  which we associate with  the decrease of the superfluid fraction of the cloud. In addition, we plot with solid lines the result of a simulation of the experimental protocol with the GPE. We observe an excellent agreement for both $c_x$ and $c_y$. The superfluid fraction $f_{s,x}$ obtained from the ratio \eqref{fscxcy} is plotted in Fig.\,\ref{fig:fsmeasured}.

\textit{Discussion and conclusion.} The determination of the superfluid fraction $f_{s,x}$ based on sound propagation is in excellent agreement with the prediction from the GPE. The determination of $f_{s,x}$ based on Leggett's formula, although limited by the finite resolution of our optical system, also agrees  well with the prediction.

More generally, one may favor one of the two methods depending on the system under study. For example, in a spin-orbit coupled BEC that violates Galilean invariance \cite{zhang2016superfluid}, the sound velocity measurement will give access to the superfluid fraction, while the density may remain uniform, in which case Leggett's bound is not relevant. Conversely in supersolid BECs \cite{leonard2017supersolid, Li17,Chomaz19,Bottcher19,Tanzi19,Norcia21,Tanzi21}, the excitation spectrum is, in general, more complex \cite{josserand2007patterns, Hofmann21} and the sound velocity measurement is not directly applicable to extract $f_s$, which, at least in  one-dimensional-like configurations, may be instead calculated using  Leggett’s formula \cite{Roccuzzo21}. A challenging question concerns  the determination of the superfluid density in higher dimensions if  the total density profile is not factorizable  along the various directions \cite{Leggett1998}, as in the case of 2D dipolar supersolids \cite{Norcia21} and of a vortex lattice \cite{Sonin87}. Our work also paves the way for the investigation of the superfluid fraction in other density modulated quantum gases, like Fermi superfluids and 1D and disordered systems. It could be extended to study the links between the quenching of the superfluid fraction and the emergence of number squeezing and phase fluctuations effects \cite{Orzel01,smerzi2002dynamical} for deep periodic potentials as well as to investigate the consequence of finite temperature effects.

While we were completing this work, a manuscript by Tao et al. explored anisotropic superfluidity in a periodically modulated trapped BEC gas \cite{Tao2023}. We employ here lattices with much larger periods, allowing for an explicit measurement of Leggett's formula and enhancing the role of two-body interactions.

\begin{acknowledgments}
\textit{Acknowledgments.} 
G.C., C.M. and F.R. contributed equally to this work. We acknowledge the support by ERC (Grant Agreement No 863880) and ANR (Grant Agreement ANR-18-CE30-0010). We thank Brice Bakkali-Hassani for his participation at the early stage of the project. We also acknowledge Tony Leggett, Anne-Laure Dalibard, Stefano Giorgini, Augusto Smerzi, Alessio Recati, Christopher Pethick, and Ian Spielman for fruitful discussions.
\end{acknowledgments}

\bibliography{bibliography}

\begin{thebibliography}{58}
\bibitem [{REF()}]{REFSM}%
  \BibitemOpen
  \href@noop {} {\enquote {\bibinfo {title} {For details, see {S}upplemental
  {M}aterials, which includes {R}efs.\,\cite{Dorrer07,Corman17}},}\
  }\BibitemShut {NoStop}%
\end{thebibliography}

\section{Supplemental Material}

\subsection{Leggett's bound and Gross-Pitaevskii equation}

We consider a weakly interacting Bose-Einstein condensate (BEC), described by Gross-Pitaevskii theory with the order parameter $\psi(\bs r,t)$ and the density $\rho(\bs r,t)=|\psi(\bs r,t)|^2$. We address the case investigated in the main text, where the BEC is confined in a box of length $L$ in the presence of an external potential $V(x)$. The equilibrium density $\rhoe(x)$ thus depends  only on the variable $x$. We assume that the potential $V(x)$ satisfies the periodicity  condition $V(L)=V(0)$ at the boundary of the box.  

We provide below a direct proof that Leggett's upper bound 
\begin{equation}
    f_{s,x}  \le \left(\langle \rhoe\rangle\; \langle \frac{1}{\rhoe}\rangle \right)^{-1}
    \label{leggettS}
\end{equation}
for the superfluid fraction, relative to a fluid moving along the $x$-direction, reduces to an identity. Here the average value of a quantity $f(x)$ is defined by
\begin{equation}
\langle f\rangle =\frac{1}{L}\int_0^L f(x)\,dx.
\label{eq:def_average}
\end{equation}

In the main text, we take $V(x)=V_0\cos(qx)$ which is spatially periodic with period $\ell=2\pi/q$. In this case, the density $\rho_{\rm eq}(x)$ is also periodic with period $\ell$. Then, provided there is an integer number of periods $\ell$ in the box of length $L$, the definition of the average value (\ref{eq:def_average}) for a periodic function $f(x)$ coincides with the definition of the main text
\begin{equation}
\langle f\rangle =\frac{1}{\ell}\int_0^\ell f(x)\,dx.
\end{equation}

In the frame moving with velocity $v_0$, \emph{i.e.,} in the presence of the perturbation $-v_0P_x$ (see main text), the Gross-Pitaevskii equation (GPE) gives rise to the following expression for the equation of continuity (only the motion along the $x$-axis is considered here)
\begin{equation}
	\partial_t \rho(x,t) + \partial_x \left\{\rho(x,t)\left[v(x,t)-v_0\right]\right\} = 0 \; ,
	\label{continuity}
\end{equation}
where $v(x)= (\hbar/m)\partial_x \phi(x)$ is the superfluid velocity field, fixed by the gradient of the phase of the order parameter. In the limit of small $v_0$, one can replace the density $\rho(x,t)$ with the equilibrium value $\rhoe(x)$.

We look for a stationary solution in the moving frame by imposing a stationary flow $\rhoe(x)(v(x)-v_0)=j$, corresponding to a position and time independent current $j$. We then obtain the result 
\begin{equation} 
	\frac{\hbar}{m}\partial_x\phi(x) = v(x)=v_0 +\frac{j}{\rhoe(x)} 
	\label{v(x)} 
\end{equation} 
for the gradient of the phase of the order parameter. Imposing  that the phase satisfies the periodic boundary condition $\phi(L)=\phi(0)$, the integration of Eq.(\ref{v(x)}) gives
\begin{equation}
0=v_0L+j\int_0^L \frac{dx}{\rhoe(x)}
\end{equation}
giving the current
\begin{equation}
  j=-\frac{v_0}{\langle \frac{1}{\rhoe}\rangle}.
	\label{j} 
\end{equation} 
Finally, we calculate the average value of the momentum operator using Eqs.\,(\ref{v(x)},\ref{j}):
\begin{eqnarray}
\langle \hat P_x\rangle &=& m\int \rhoe(x)\,v(x)\,dx 
= mL \left( v_0 \langle \rhoe\rangle +j  \right) \nonumber\\
&=& Nmv_0 \left(1-\frac{1}{\langle \rhoe\rangle \;\langle \frac{1}{\rhoe}\rangle}  \right)
\label{}
\end{eqnarray}
where we introduced the number of particle $N=L\langle \rhoe\rangle$.
Using the definition  of the main text:
\begin{equation}
f_{s,x}=1-\lim_{v_0\to 0}\frac{\langle \hat P_x\rangle }{Nmv_0},
\end{equation}
we straightforwardly obtain the announced result for the superfluid fraction:
\begin{equation}
    f_{s,x}  = \left(\langle \rhoe\rangle\; \langle \frac{1}{\rhoe}\rangle \right)^{-1}.
    \label{Leggett_equality}
\end{equation}

\subsection{Superfluid density and effective mass}

In this section, we explain why the two seemingly unconnected quantities,  superfluid fraction and  effective mass, are connected when one considers a zero-temperature Bose gas placed in a periodic potential and described by the GPE for the order parameter $\psi(x)$. We assume that the potentiel $V(x)$ is spatially periodic with period $\ell$ and that there are an integer number of periods $n_p=L/\ell$ in the box of length $L$.

\paragraph{Superfluid density.}
As explained in the previous section, the superfluid density can be calculated using linear response theory for the perturbation $-v_0\hat P_x$. Equivalently, one can apply twisted boundary conditions along the $x$ axis, $\psi(L)=e^{i\theta}\,\psi(0)$ (with $\theta$ arbitrary small), and look for the increase of energy of the ground state $\psi(x)$ at second order in $\theta$ \cite{Leggett1970}. This energy increase directly provides  $f_{s,x}$:
\begin{equation}
E(\theta)\approx E(0)+N f_{s,x}\frac{\hbar^2\theta^2}{2mL^2}.
\label{eq:E_twisted}
\end{equation}
Since the potential $V(x)$ is periodic, it is natural to assume that the phase twist $\theta$ of the wave function $\psi(x)$ minimizing the energy is uniformly distributed over all lattice periods, hence
\begin{equation}
\psi\left(n\ell\right)=e^{i\theta/n_p}\,\psi\left((n-1)\ell\right)\qquad n=1,\cdots,n_p.
\label{eq:BC_simplified}
\end{equation} 

To determine the ground state $\psi(x)$, it is convenient to perform the gauge transform $\phi(x)=e^{-i \theta x/L}\,\psi(x)$, so that we recover periodic boundary conditions for $\phi(x)$ with an additional vector potential. More precisely, $\phi(x)$ is the solution of the equation
\begin{equation}
\frac{\left(\hat P_x+\hbar\theta/L\right)^2}{2m}\phi(x) +V(x)\phi(x)+g|\phi(x)|^2 \phi(x)=\mu \phi(x)
\label{eq:TwBC}
\end{equation} 
minimizing the GP energy functional with the boundary condition deduced from Eq.\,(\ref{eq:BC_simplified}):
\begin{equation}
\phi\left(n\ell\right)=\phi\left((n-1)\ell\right),
\label{eq:BC_phi}
\end{equation} 
\emph{i.e.,} a periodic boundary condition over each lattice period. 
This procedure is directly connected to the one of the previous section by setting $v_0=\hbar\theta/mL$ and leads to the result (\ref{Leggett_equality}).

\paragraph{Effective mass.} We now turn to the determination of the effective mass for a BEC placed in the periodic potential $V(x)$. We look for solutions that can be written as Bloch functions, $\psi_k(x)=e^{i kx}\,u_k(x)$, where $u_k(x)$ is periodic over the lattice and satisfies 
\begin{equation}
u_k\left(n\ell\right)=u_k\left((n-1)\ell\right).
\label{eq:BC_uq}
\end{equation}
The equation satisfied by $u_k(x)$ is
\begin{equation}
\frac{(\hat P_x+\hbar k)^2}{2m}u_k(x)+V(x)u_k(x)+ g|u_k(x)|^2 u_k(x)=\mu\, u_k(x)
\label{eq:eff_mass}
\end{equation}
and the effective mass $m_x^*$ is defined by the curvature of the energy around $k=0$: 
\begin{equation}
E(k)\approx E(0)+N\frac{\hbar^2 k^2}{2m_x^*}.
\label{eq:bottom_band}
\end{equation}
Note that  we define here the effective mass using the solutions of the non-linear GPE. One can show that the same value for $m^*_x$ arises if one considers the Bogoliubov excitations of the gas in the periodic potential $V(x)$, i.e. the linearized version of the problem (see Eq.\,(28) of \cite{taylor2003bogoliubov}).

The mathematical similarity between the two problems (\ref{eq:TwBC}-\ref{eq:BC_phi}) and (\ref{eq:BC_uq}-\ref{eq:eff_mass}) is clear, with the replacement $\theta/L \leftrightarrow k$, and the comparison between Eqs.\,(\ref{eq:E_twisted}) and (\ref{eq:bottom_band}) provides the relation 
\begin{equation}
m_x^*=m/f_{s,x}
\end{equation} for the zero-temperature Bose gas described by the GPE.

\begin{figure*}[t!!!]
\begin{center}
\includegraphics[height=3.5cm]{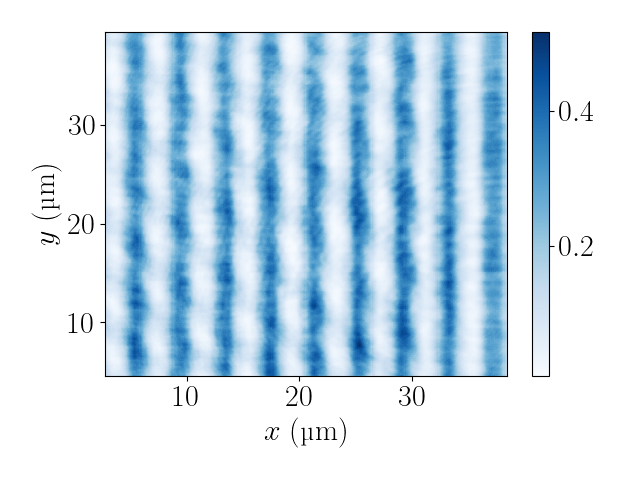}
\includegraphics[height=3.5cm]{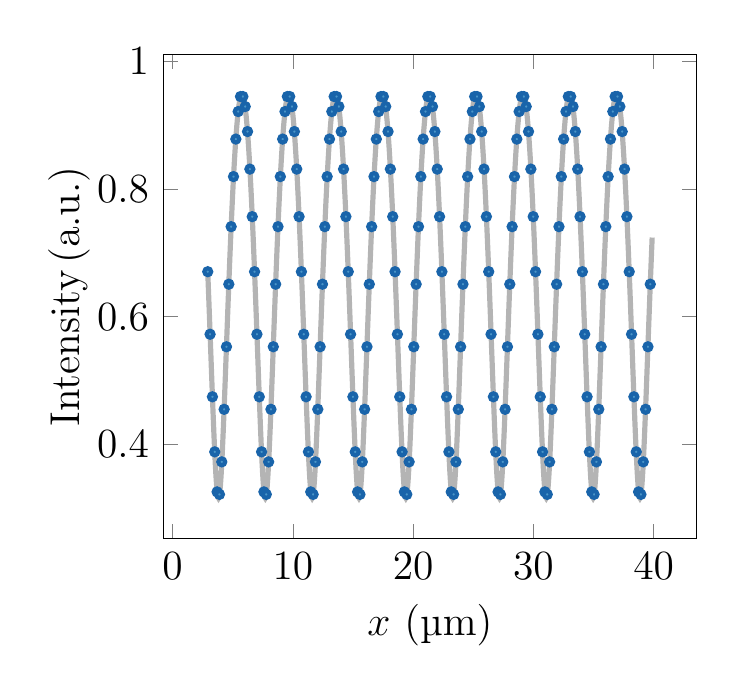}
\includegraphics[height=3.5cm]{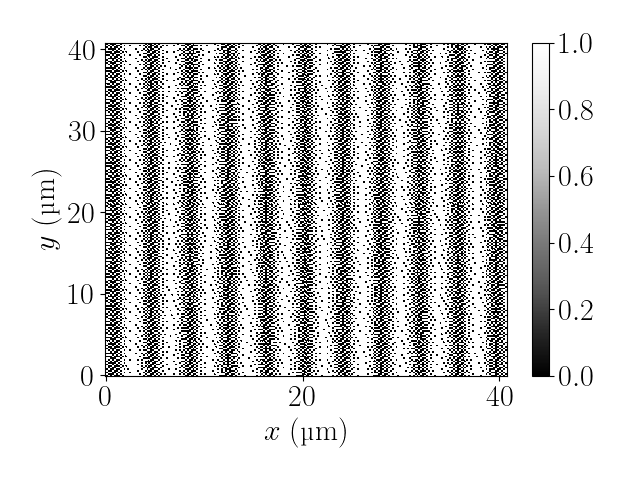}
\includegraphics[height=3.5cm]{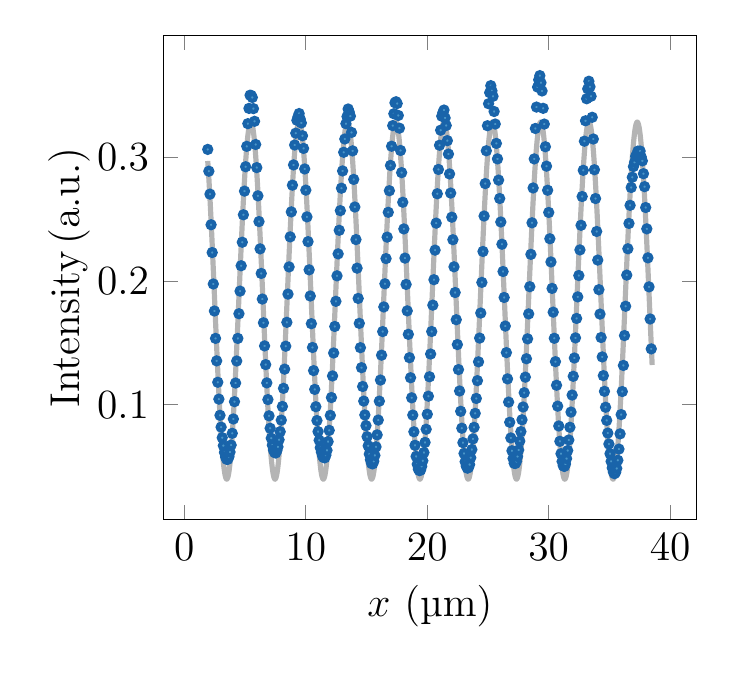}
\end{center}
\vskip-120pt
\hskip-80pt(a) \hskip120pt (b)\hskip100pt (c)\hskip100pt (d)
\vskip90pt
\caption{(a) Typical dithered pattern displayed on the DMD. Each pixel is either white (1) or black (0). (b) Cut of the dithered pattern (circles) along with a fit (solid line) of the data to the function $x \mapsto\sqrt{a_0+ a_1\cos\left(kx\right)}$. (c) Measured image of the projected lattice on the auxiliary camera. The residual intensity modulation along $y$ is an artifact of the optical measurement due to parasitic fringes. The axes are rescaled to coincide with the effective position in the atoms plane. (d) Cut of the auxiliary camera image along the $x-$axis along with a fit to the function  $x\mapsto b_0 + b_1 \cos\left(kx\right)$.}\label{fig:latticeDMD}
\end{figure*}

\subsection{Preparation and excitation of the gas}
Our experimental setup has been described in Ref.\,\cite{Ville17}. We load an optical box potential made of a combination of laser beams at a wavelength of 532\,nm. The confinement along the vertical direction $z$ is created by an optical lattice and we load a 3D Bose-Einstein condensate in a single node of this lattice. The resulting trapping potential is well approximated by an harmonic potential of frequency $\omega_z\approx2\pi \times 3.7\,$kHz. The in-plane confinement is a square box potential of size $L=40\,\mum$ obtained by imaging the chip of a digital micro-mirror device (DMD) on the atomic plane with a high-resolution microscope objective. Atoms are cooled down thanks to evaporative cooling to temperatures  $T\lesssim20\,$nK. For all experiments reported in this work,  the average surface density is fixed to $\rho_0=60(3)$ atoms$/\mum^2$. Atoms are polarized in the electronic ground state $\ket{F=1,m=-1}$. The in-situ density distribution is measured using partial transfer to the $\ket{F=2}$ state, followed by absorption imaging onto a CCD camera with a typical resolution of $\lesssim 1\,\mum$. The method of partial transfer allows us to control the amount of atoms sensitive to the imaging beam so as to maintain the optical depth of the imaged cloud to a low value. In this regime, we avoid any collective effects in the light diffusion by the atomic cloud \cite{Corman17}.

The excitation of sound waves is performed by applying a magnetic field gradient over the size of cloud, which creates a uniform force along either the $x$ or the $y$ direction. In this work, we  use a gradient of $b'=3.4\,\mu$G/$\mum$. For a chemical potential of $\mu_0/k_B \sim  50\,$nK of the cloud, this corresponds to a peak-to-peak potential of 0.09\,$\mu_0$. The speed of sound is measured by suddenly removing this gradient at time $t=0$ and then monitoring the time evolution of the center-of-mass distribution of the cloud. The typical amplitude of oscillation of the center-of-mass along the axis of the excitation is $\approx 1\,\mum$ and no clear oscillation signal is observed along the orthogonal direction.

\subsection{Optical lattice potential}
We describe in this paragraph the preparation and the characterization of the sinusoidal optical potential applied to the atomic cloud. This potential is created by imaging the chip of a DMD onto the atomic plane with an imaging system of resolution $\lesssim 1\,\mum$. The pixel size of the DMD is 13.7$\,\mum$ and the magnification is 1/70 leading to an effective pixel size on the atomic cloud of $\approx\,$0.2$\,\mum$, much smaller than the optical resolution of the imaging system. The laser wavelength is 532\,nm and the beam waist in the atomic plane is $\sim 85\,\mum$ so that the intensity profile over the atomic cloud size is almost uniform.

As described in Ref.\,\cite{Zou21}, we use a dithering algorithm to create effective ``grey levels" of light intensity on the atomic plane \footnote{We bypass the feedback loop technique reported in Ref.\,\cite{Zou21}, which relies on LDA.}. The lattice optical potential experienced by the atoms is measured with an auxiliary imaging system with a similar resolution $\lesssim 1\,\mum$ and a magnification $\sim 3.2$. As the DMD operates as an amplitude modulator \cite{Dorrer07}, a sinusoidal modulation of the intensity on the atom plane will be obtained using a target profile for the dithering algorithm given by
\begin{equation}
f(x)=\sqrt{\alpha+\beta \cos (kx)}.
\end{equation}
The coefficients $\alpha$ and $\beta$ are chosen so that $\forall x$, $0<f(x)$. For the data reported in this work, obtained with a lattice of period $d\approx4\,\mum$, we used  $\alpha=0.5$ and $\beta=0.4$.

We show in Fig.\,\ref{fig:latticeDMD} a typical example of a dithered image that we display on the DMD. We also report the corresponding intensity distribution over the size of the cloud measured with the auxiliary imaging system. The obtained lattices are well sinusoidal with an almost constant amplitude over the size of the cloud. We also checked that the maximal amplitude of the lattice is linear with the laser light intensity.

We now discuss the calibration of the lattice depth $V_0$. We shone  large-period lattices ($d> 8\,\mum$) on the atomic cloud, we determined the dominant term of the density modulation $\rho_1$ and we deduced $V_0$ using the LDA prediction $\rho_1=V_0/g$. To make this approach relevant, we restrict ourselves to low lattice depths $V_0< \mu_0$  and we checked that, in this range and for a fixed laser intensity, the contrast of the density modulation $\mathcal{C}=\rho_1/\rho_0$ is independent of the lattice period. For such large periods, our finite imaging resolution does not significantly affect the measured signal, contrary to the case of shorter periods. Its measured value, $\mathcal C\approx 0.75$, is consistent with the chosen values of $\alpha$ and $\beta$, leading to $\mathcal C=0.8$. Note that for large-period lattices ($\epsilon_q\ll\mu_0$), the correction to the  LDA discussed in the main text (see Eq.\,8) is also negligible.

\subsection{Role of finite optical resolution on the computation of Leggett's formula}
For strong enough lattice depths, $V_0 \gtrsim \mu_0$, the density modulation $\rho_n(x)$ differs from a sinusoidal profile. While the determination of the superfluid fraction through Leggett's formula is still exact (in the regime where the GPE is valid), its experimental determination becomes more challenging. Deviations to a sinusoidal profile are described by the emergence of higher-order harmonics $\rho_n$ in the density modulation, which we have defined through   $\rho_0-\sum_{n>0}\rho_n\cos(nqx)$. These harmonics are subjected to spatial filtering by our imaging system. This filtering is modeled by the $\beta_n$ coefficients in the main text. 

\begin{figure*}[ht!]
\begin{flushleft}
(a)\hskip160pt (b)\hskip160pt (c)
\end{flushleft}
\begin{center}
\vskip-16pt
\includegraphics[height=4cm]{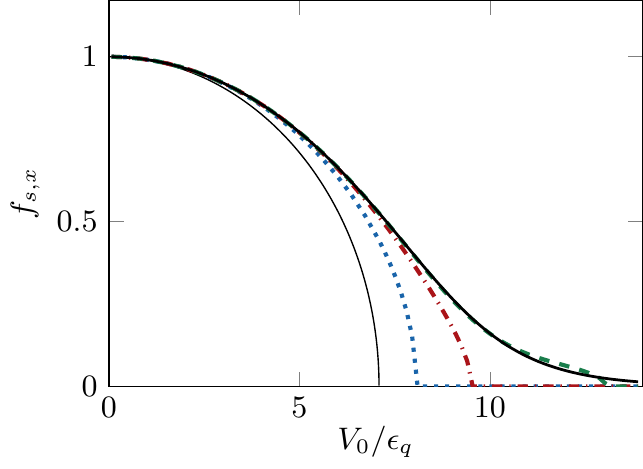}\hskip8pt
\includegraphics[height=4cm]{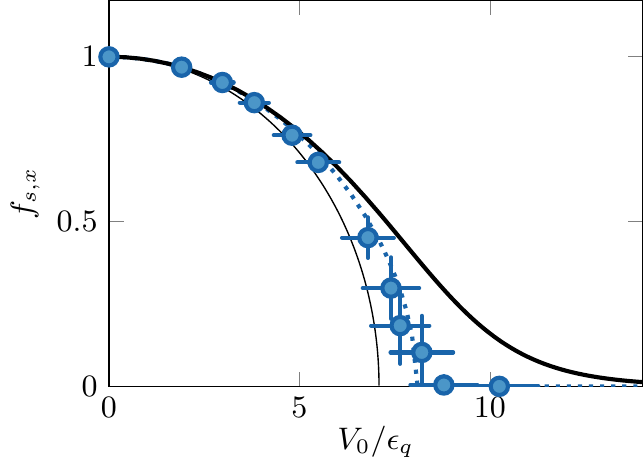}\hskip8pt
\includegraphics[height=4cm]{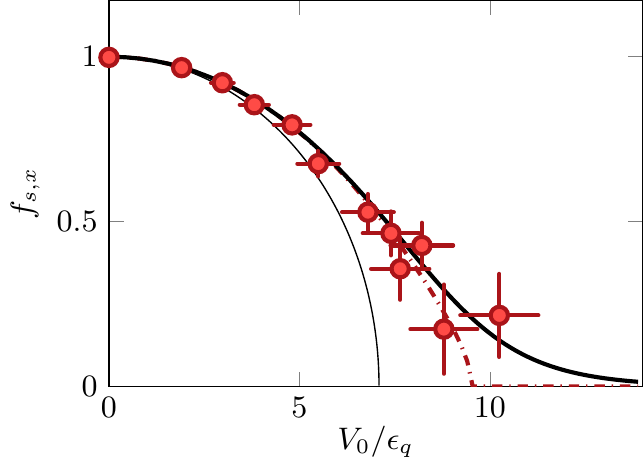}
\end{center}
\caption{Influence of the finite optical resolution on the determination of the superfluid fraction using Leggett's formula. For all plots, we show the LDA (thin solid line) and the GPE (thick solid line) predictions. In (a), the predictions of the GPE with a filtering of the lowest harmonics are shown when keeping only the first harmonic (blue dotted line), the two first harmonics (red dot-dashed line) and the three first harmonics (green dashed line). (b) Experimental determination of the Leggett's formula keeping only the first harmonic (blue circles). (c) Same when keeping only the two first harmonics, as shown in the main text (red circles). }
\label{fig:GPfilter} 
\end{figure*}

We show in Fig.\,\ref{fig:GPfilter}(a) the value of Leggett's formula computed from the ground state of the GPE taking into account this spatial filtering. In addition to the LDA (thin solid line) and the GPE predictions (thick solid line), we show the results obtained from the GPE  keeping only the first $n=1$ harmonic (dotted line), the two first harmonics (dot-dashed line) and the three first harmonics (dashed line). With only two harmonics of the density modulation, the estimated superfluid fraction is very close to  the exact prediction for $f_{s,x}\gtrsim 0.3$ and slightly deviates for lower superfluid fractions. Including the third harmonic leads to an excellent approximation down to  $f_{s,x}\sim 0.1$. In Fig.\,\ref{fig:GPfilter}(b), we compare  our measurement of the density modulation including only the first harmonic together with the corresponding prediction of the GPE with filtering. The agreement is excellent. We also note that for $V_0/\epsilon_q \gtrsim8$ the computed superfluid fraction is null because $\rho_1>\rho_0$, which  leads to a vanishing  density $\rho_0-\rho_1\cos(qx)$ around the lattice maxima. Finally, we show in Fig.\,\ref{fig:GPfilter}(c) the data reported in the main text and compare them with the prediction of the GPE when keeping the two first harmonics ($\rho_1$ and $\rho_2$). Here again the agreement is very satisfactory. Note that the observation of the third harmonic is not accessible with our setup. It corresponds to a density modulation  with a period $\sim 1.3\,\mum$. To be detected, it would require a numerical aperture (NA) of the optical system around 0.6 and is thus fully filtered for our NA of 0.45. 
These results show that our choice of parameters, and especially the lattice period,  is suitable to obtain, within our experimental uncertainties, a robust estimate of the superfluid fraction with Leggett's formula in a broad range of values of the lattice depth. 

\end{document}